\documentclass[twocolumn]{jpsj3}

\usepackage{amsmath}
\usepackage{amsfonts}
\usepackage{amssymb}
\usepackage{cite}
\usepackage{txfonts}
\usepackage{bm}
\usepackage{tabularx}
\usepackage{graphicx,color}

\def\Hc2{H_\mathrm{c2}}
\def\Tc{T_\mathrm{c}}

\author{
	Shunichiro \textsc{Kittaka}$^{1,\thanks{E-mail: kittaka@issp.u-tokyo.ac.jp}}$, 
	Koji \textsc{An}$^{1}$,
	Toshiro \textsc{Sakakibara}$^{1}$, \\
	Yoshinori \textsc{Haga}$^{2}$,
	Etsuji \textsc{Yamamoto}$^{2}$,
	Noriaki \textsc{Kimura}$^{3}$, 
	Yoshichika \textsc{\={O}nuki}$^{2,4}$, and
	Kazushige \textsc{Machida}$^{5}$
}
\inst{$^{1}$Institute for Solid State Physics, University of Tokyo, Kashiwa, Chiba 277-8581, Japan\\
      $^{2}$Advanced Science Research Center, Japan Atomic Energy Agency, Tokai, Ibaraki 319-1195, Japan\\
      $^{3}$Department of Physics, Tohoku University, Sendai 980-8578, Japan\\
      $^{4}$Department of Physics, Osaka University, Toyonaka, Osaka 560-0043, Japan\\
      $^{5}$Department of Physics, Okayama University, Okayama 700-8530, Japan
}

\begin{document}

\title{Anomalous Field-Angle Dependence of the Specific Heat of Heavy-Fermion Superconductor UPt$_3$}

\date{\today}

\abst{
We have investigated the field-angle variation of the specific heat $C(H,\phi,\theta)$ of the heavy-fermion superconductor UPt$_3$ at low temperatures $T$ down to 50~mK,
where $\phi$ and $\theta$ denote the azimuthal and polar angles of the magnetic field $H$, respectively.
For $T=88$~mK, $C(H,\theta=90^\circ)$ increases proportionally to $\sqrt{H}$ up to nearly the upper critical field $\Hc2$,
indicating the presence of line nodes.
By contrast, $C(H,\theta=0^\circ)$ deviates upward from the $\sqrt{H}$ dependence for $\sqrt{H/\Hc2} \gtrsim 0.5$.
This behavior can be related to the suppression of $\Hc2$ along the $c$ direction,
whose origin has not been resolved yet.
Our data show that the unusual $\Hc2$ limit becomes marked 
only when $\theta$ is smaller than $30^\circ$.
In order to explore the possible vertical line nodes in the gap structure,
we measured the $\phi$ dependence of $C$ in wide $T$ and $H$ ranges.
However, we did not observe any in-plane angular oscillation of $C$ within the accuracy of $\Delta C/C \sim 0.5\%$.
This result implies that field-induced excitations of the heavy quasiparticles occur isotropically with respect to $\phi$,
which is apparently contrary to the recent finding of a twofold thermal-conductivity $\kappa(\phi)$ oscillation.
}

\kword{UPt$_3$, spin-triplet superconductivity, superconducting gap, upper critical field, specific heat}

\maketitle

\section{Introduction}
\begin{figure}
\begin{center}
\includegraphics[width=3.2in]{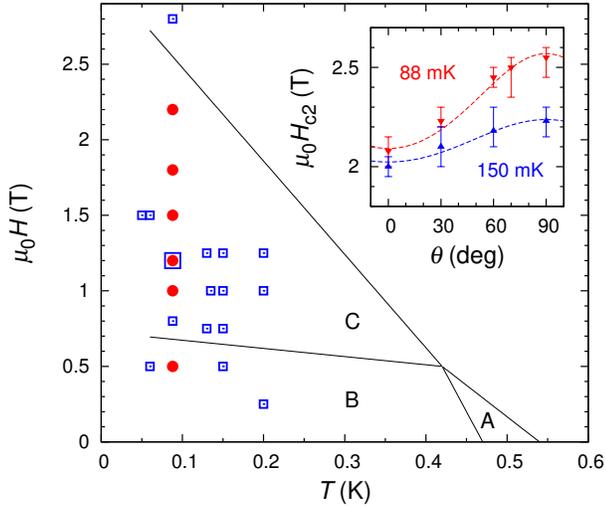}
\end{center}
\caption{
(Color Online) Schematic field-temperature phase diagram of UPt$_3$ for $H \parallel ab$. 
Squares (circles) represent the field and temperature at which 
$C(\phi)$ measurements are performed with $\theta=90^\circ$ ($60^\circ$).
The inset represents the $\theta$ dependence of $\Hc2$ determined from the $C(H)$ data obtained at 88 and 150~mK.
Dashed lines are guides to the eye.
}
\label{HT}
\end{figure}

UPt$_3$ is an unconventional superconductor with multiple superconducting (SC) phases \cite{Joynt2002RMP} (A, B, and C phases, see Fig.~\ref{HT}).
In particular, strong evidence for spin-triplet superconductivity has been provided by NMR Knight shift measurements,\cite{Tou1998PRL,Machida1998JPSJ}
and the presence of line nodes in the SC gap has been clearly indicated by the power-law behavior in thermodynamic and transport quantities. \cite{Shivaram1986PRL,Kohri1988JPSJ,Broholm1990PRL,Ramirez1995PRL,Suderow1997JLTP}
To explain the multiple SC phases, the presence of a two-component order parameter coupled to a symmetry-breaking field (SBF) has been proposed;
whereas two order parameters are degenerate in the B phase, one of them appears in the A phase and the other appears in the C phase. 
Although the SBF, which lifts the degeneracy of the order parameters at zero field, is still controversial,
it could possibly be related to short-range antiferromagnetic (AFM) fluctuations in the basal plane developing below 5~K~\cite{Aeppli1989PRL,Hayden1992PRB,Machida1995JPSJ}
and/or distortions in the crystal structure~\cite{Midgley1993PRL,Mineev1993JETPL,Walko2001PRB}.

While UPt$_3$ has been well established to have nodes in the SC gap, 
the precise gap structure of each SC phase has been debated actively.
On theoretical grounds, 
various multi-component order parameters have been proposed so far.~\cite{Joynt2002RMP,Sauls1994JLTP,Machida1999JPSJ}
One of the plausible scenarios is the $E_{2u}$ model with spin-triplet pairing involving 
the SC order parameters $k_z(k_x^2-k_y^2)$ and $2k_xk_yk_z$, 
whose gap functions have \textit{fourfold} symmetry with vertical line nodes and are degenerate in the B phase as $|k_z|(k_x^2+k_y^2)$ with axial symmetry.
Experimental evidence supporting this scenario has been provided by 
small-angle neutron scattering \cite{Huxley2000Nature,Champel2001PRL} and junction experiments \cite{Strand2009PRL,Strand2010Science}
performed in the A and B phases.
However, a conflicting scenario, the spin-triplet $E_{1u}$ model with the order parameters $k_x(5k_z^2-1)$ and $k_y(5k_z^2-1)$ has recently been proposed 
on the basis of the results of field-angle-resolved thermal-conductivity $\kappa(\phi)$ measurements performed in the B and C phases;\cite{Machida2012PRL,Tsutsumi2012JPSJ}
under a magnetic field $H$ rotated around the $c$ axis, $\kappa(\phi)$ exhibits a \textit{twofold} oscillation at 50~mK only in the C phase.

In addition to the SC gap structure, 
the origin of the strong limit of the upper critical field $\Hc2$ for $H \parallel c$ at low $T$ has not been resolved.
Although it resembles the $\Hc2$ limit due to the Pauli paramagnetic effect, \cite{Tenya1996PRL,Choi1991PRL} 
$\Hc2$ cannot be affected by this effect in UPt$_3$
because the results of NMR Knight shift experiments suggested that the $d$ vector of the triplet Cooper pair is perpendicular to $H$ in the C phase;
the Cooper pairs can be polarized along the field direction. \cite{Tou1998PRL}
This is reminiscent of a similar, unusual $\Hc2$ limit observed in another plausible spin-triplet superconductor Sr$_2$RuO$_4$ for $H \parallel ab$.\cite{Kittaka2009PRB, Maeno2012JPSJ}

To obtain hints for clarifying the precise gap structure and the origin of the $\Hc2$ limit in UPt$_3$, 
we have performed field-angle-resolved specific-heat $C(H,\phi,\theta)$ measurements down to 50~mK,
where $\phi$ is the azimuthal field angle within the $ab$ plane measured from the crystalline [1$\bar{2}$10] axis and $\theta$ denotes the polar angle between $H$ and the $c$ axis.
Because $C/T$ is proportional to the quasiparticle (QP) density of states (DOS) at low $T$,
this technique enables us to get an insight into nodal directions of bulk superconductors
from the field-orientation dependence of QP excitations.\cite{Sakakibara2007JPSJ}
At 88~mK, an unusual increase in $C(H)$ for $\sqrt{H/\Hc2} \gtrsim 0.5$ was found only when $|\theta| \le 30^\circ$.
Correspondingly, the $\theta$ dependence of $C$ at 88~mK exhibits a peak at $\theta=0^\circ$ with a width of $|\theta| \lesssim 30^\circ$ in the high-$H$ regime, 
whereas $C(\theta)$ at low $H$ shows a twofold oscillation with a deep minimum at $\theta=0^\circ$. 
This feature indicates the occurrence of unusual QP excitations for $|\theta| \lesssim 30^\circ$ in the high-$H$ regime,
which might be related to the $\Hc2$ limit.
Under $H$ rotated around the $c$ axis, we observed no variation in $C(\phi)$ down to 50~mK ($\sim 0.1\Tc$),
implying a small change in QP DOS as a function of in-plane field angle.

\section{Experimental Details}
A single crystal of UPt$_3$ used in the present study ($\Tc=0.52$ K, 54.2 mg) was grown by the Czochralski pulling method.
The sample was shaped into a disk with flat surfaces parallel to the $ab$ plane.
The specific heat $C$ was measured by relaxation and standard quasi-adiabatic heat-pulse methods in a dilution refrigerator (Oxford Kelvinox AST Minisorb).
In all the data presented, a nuclear Schottky contribution ($C_\mathrm{n}$) by the Zeeman splitting of $^{195}$Pt nuclei ($I=1/2$) 
and the addenda contribution were subtracted.
It was confirmed that the addenda specific heat is always less than 4\% of the sample specific heat and exhibits a negligible field-angle dependence.
Although a rapid increase was observed in $C/T$ at zero field below 0.1~K, \cite{Brison1994JLTP,Kittaka2012JPCS}
it is probably related to an AFM long-range ordering at 20~mK~\cite{Schuberth1992PRL,Koike1998JPSJ}.
Magnetic fields were generated in the $xz$ plane using a vector magnet 
composed of horizontal split-pair (5~T) and vertical solenoidal (3~T) coils.
By rotating the refrigerator around the $z$ axis using a stepper motor mounted at the top of a magnet Dewar,
we controlled the field direction three-dimensionally.
Thus, we performed experiments in an accurate ($< 0.1^\circ$) and precise ($< 0.01^\circ$) field alignment with respect to the $ab$ plane,
which was adjusted by using the anisotropy of the specific heat of UPt$_3$ in the $ac$ plane.
For each point, the specific-heat value was determined by the average of about ten successive measurements.

\section{Results and Discussion}
\subsection{Field dependence of the specific heat}
\begin{figure}
\begin{center}
\includegraphics[width=3.2in]{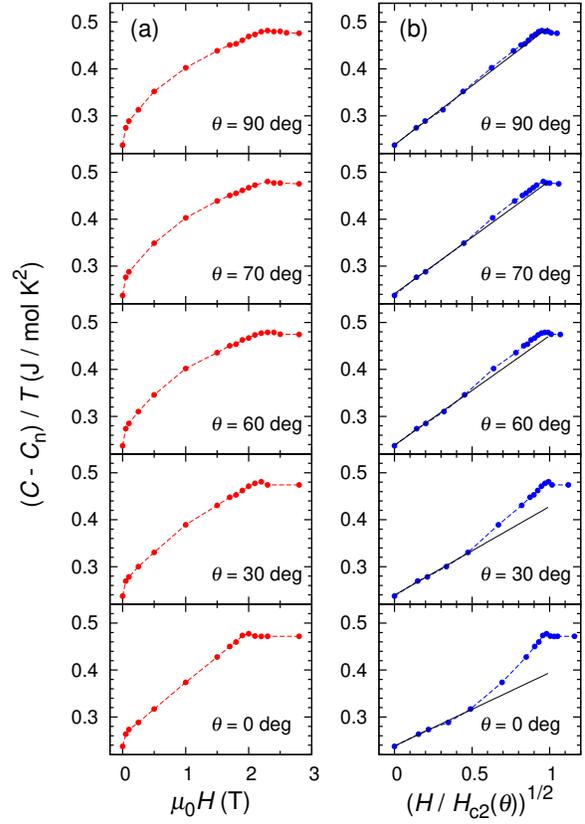}
\end{center}
\caption{
(Color Online) (a) Magnetic-field dependence of $(C-C_\mathrm{n})/T$ at several polar angle $\theta$ values at 88~mK.
(b) $(C-C_\mathrm{n})/T$ as a function of $\sqrt{H/\Hc2(\theta)}$.
Dashed lines are guides to the eye.
Solid lines represent the fits to the data using the function $a\sqrt{H}+b$ in the low-field regime ($\mu_0H \le 0.5$~T).
}
\label{Hdep}
\end{figure}

Figure~\ref{Hdep}(a) shows the $H$ dependence of the nuclear-subtracted $C$ divided by $T$ at various $\theta$ values.
These data were obtained at 88~mK and $\phi=0^\circ$.
At low $H$ below 0.5~T, $C(H)$ increases rapidly with $H$ at any $\theta$.
This rapid increase is fitted well using the function $a\sqrt{H}+b$, 
as represented by solid lines in Fig.~\ref{Hdep}(b), 
where $C/T$ is plotted as a function of the square root of $H/\Hc2(\theta)$.
The occurrence of the $\sqrt{H}$ behavior at any $\theta$ indicates that 
Doppler-shifted QP excitations around line nodes are predominant in determining $C(H)$ and
excludes the possibility of forming a SC gap structure with only point nodes at the poles.~\cite{Yano2008PRL}

If $\Hc2$ is dominated by the orbital pair-breaking effect, 
the $C(H)$ of anisotropic superconductors with line nodes is expected to exhibit the $\sqrt{H}$-linear dependence up to nearly $\Hc2$.\cite{Ichioka2007PRB}
Such behavior is indeed observed in $C(H)$ at $\theta=90^\circ$.
By contrast, at $\theta=0^\circ$, $C$ as a function of $\sqrt{H}$ exhibits a gradual upturn in the high-$H$ regime for $\sqrt{H/\Hc2} \gtrsim 0.5$, as shown in Fig.~\ref{Hdep}(b).
This upturn is clearly seen for $|\theta| \le 30^\circ$ and enhances with decreasing $\theta$ to zero.
Although the $\sqrt{H}$-linear dependence at low $H$ is slightly smeared at any $\theta$ by the effect of thermal QP excitations, 
a similar feature was observed at 150~mK as well (not shown).
The characteristic field of $0.25\Hc2$, above which the upturn becomes marked, 
is much lower than the transition field from the B phase to the C phase of about $0.7\Hc2$ for $H \parallel c$~\cite{Dijk1993JLTP}.

The field-angle $\theta$ dependence of $\Hc2$ determined from the $C(H)$ data obtained at 88 and 150~mK is shown in the inset of Fig.~\ref{HT}.
Also, a relatively strong $\Hc2$ limit occurs for $|\theta| \lesssim 30^\circ$. 
Thus, the anomalous $H$ dependence of $C(H)$ observed for $|\theta| \le 30^\circ$ seems to be related to the unresolved mechanism of the $\Hc2$ limit. 

A possible origin of the $\Hc2$ limit is the Pauli paramagnetic effect; 
the $H$ dependence of $C$ at $\theta=0^\circ$ is qualitatively similar to the calculated result based on a quasiclassical theory 
with the paramagnetic parameter $\mu=0.86$ for nodal $d$-wave superconductors.~\cite{Ichioka2007PRB}
In addition, the multi-band effect cannot be ruled out 
because UPt$_3$ has five bands in the Fermi surface \cite{McMullan2008NJP}; 
a possible change of the relevant band that determines $\Hc2$ when $T$, $H$, and field orientation are varied could lead to the apparent $\Hc2$ limit. 
Nevertheless, no strong evidence has yet been established for both possibilities. 
In particular, the Pauli effect is incompatible with the results of NMR Knight-shift experiments.\cite{Tou1996PRL,Tou1998PRL}
Thus, the origin of the $\Hc2$ limit remains unclear.

\subsection{Field-angle $\theta$ dependence of the specific heat}
\begin{figure}
\begin{center}
\includegraphics[width=3.2in]{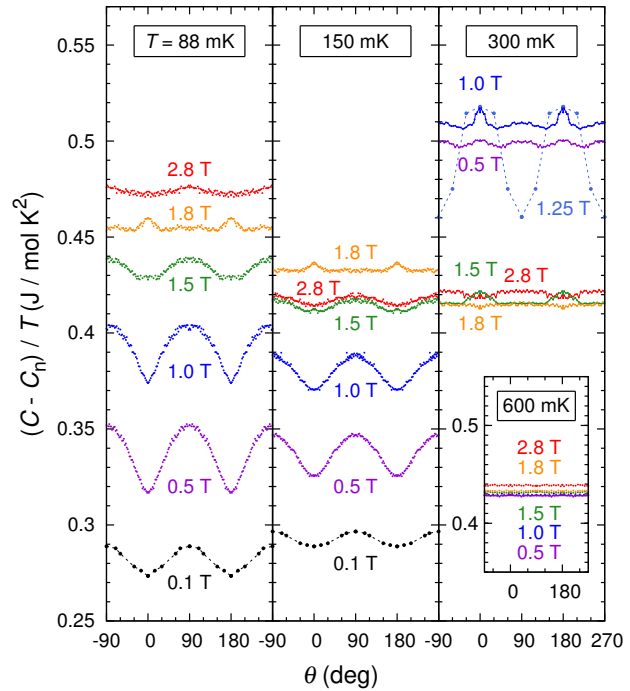}
\end{center}
\caption{
(Color Online) Polar-field-angle $\theta$ dependence of $(C-C_\mathrm{n})/T$ at several temperatures and fields.
Dashed lines are guides to the eye. For easy viewing, the data measured at the interval $0^\circ \le \theta \le 90^\circ$ are plotted repeatedly 
($\theta$ is converted to $-\theta$ and $180^\circ \pm \theta$).
}
\label{C-t}
\end{figure}

The $\theta$ dependence of $C$ measured under $H$ rotated within the $ac$ plane ($\phi=0^\circ$) is shown in Fig.~\ref{C-t},
where the $C(\theta)$ data obtained at the interval $0^\circ \le \theta \le 90^\circ$ are plotted repeatedly in accordance with the crystal symmetry.
Let us first focus on $C(\theta)$ in the normal state.
At 600~mK, which is sufficiently above $\Tc$, $C$ is invariant under the $\theta$ rotation at any $H$ below 2.8~T.
By contrast, at low $T$ below 300~mK, 
$C(\theta)$ at 2.8~T that is higher than $\Hc2$ 
exhibits a small twofold oscillation with a maximum at $\theta=90^\circ$ ($H \parallel a$). 
This anisotropy might originate from the anisotropy of the DOS enhancement by a metamagnetic transition occurring at 20~T for $H \parallel ab$.\cite{Muller1989PRB}

In the SC state, $C$ varies markedly with changing $\theta$.
At low $T$ below 150~mK and low $H$ below 1~T, $C(\theta)$ exhibits a large twofold oscillation with a deep minimum at $\theta=0^\circ$ ($H \parallel c$).
The relation $C(\theta=0^\circ) < C(90^\circ)$ means smaller QP excitations at low $H$ in $H \parallel c$ rather than in $H \parallel a$.
This anisotropy is opposite to the expectation from the $\Hc2$ anisotropy ($H_{\mathrm{c2} \parallel a} > H_{\mathrm{c2} \parallel c}$), 
which naturally leads to $C(0^\circ) > C(90^\circ)$.
This result implies that, if the $\Hc2$ limit is absent, 
the intrinsic $\Hc2(T=0~\rm{K})$ is larger for $H \parallel ab$ than for $H \parallel c$,
the same anisotropy of $\Hc2$ observed in the vicinity of $\Tc$.~\cite{Keller1994PRL}.

By increasing $H$ at 88~mK, a small peak whose width is $|\theta| \lesssim 30^\circ$ develops in $C(\theta)$ at $\theta=0^\circ$.
At 1.8~T, this peak makes the anisotropy in $C(\theta)$ reversed [$C(0^\circ) > C(90^\circ)$], 
consistent with the expectation from the observed $\Hc2$ anisotropy.
A similar peak can also be seen at 150 and 300~mK.
In particular, at 300~mK, the peak is visible even at low $H$ and with a larger width ($|\theta| \lesssim 45^\circ$).
The observed peak in $C(\theta)$ at 88~mK corresponds to the anomalous increase in $C(H)$ for $|\theta| \le 30^\circ$ shown in Fig.~\ref{Hdep}(b).
Therefore, this peak might define the conditions of the unresolved mechanism inducing the unusual $\Hc2$ limit of UPt$_3$. 

\subsection{Field-angle $\phi$ dependence of the specific heat}
\begin{figure}
\begin{center}
\includegraphics[width=3.2in]{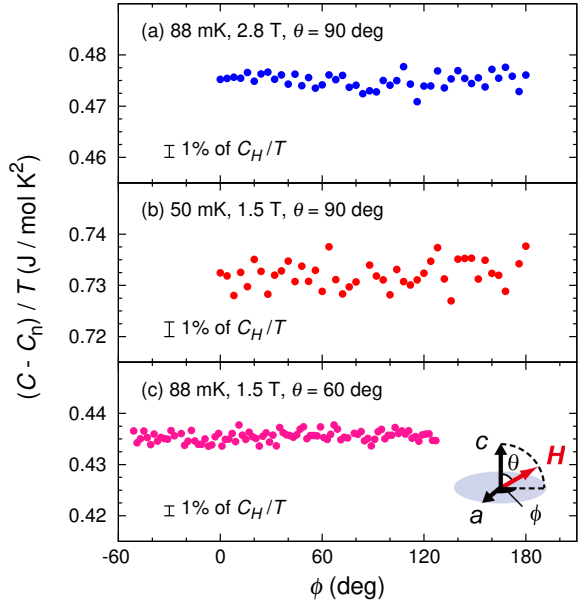}
\end{center}
\caption{
(Color Online) Azimuthal-field-angle $\phi$ dependence of $(C-C_\mathrm{n})/T$.
Bars represent 1\% of $C_H/T=[C(H)-C_{\rm n}(H)-C(0)]/T$.
}
\label{phidep}
\end{figure}

We next focus on the variation in $C$ under $H$ rotated around the $c$ axis, $C(\phi)$.
This $C(\phi)$ measurement is a powerful tool for identifying the SC gap structure.
According to the results of the Doppler-shift analysis assuming an anisotropic SC gap on a quasi-two-dimensional Fermi surface, 
zero-energy DOS, which corresponds to $C/T(\phi)$ at low $T$, is expected to be enhanced (suppressed) 
when $H$ is applied along the antinodal (nodal) direction. \cite{Vekhter1999PRB,Miranovic2003PRB,Miranovic2005JPC} 
Even at a moderate temperature, a clear $C(\phi)$ oscillation reflecting the gap structure is expected to be observed 
although its sign changes with the vortex-scattering effect.\cite{Vorontsov2006PRL,Hiragi2010JPSJ}
Indeed, a fourfold oscillation reflecting the $d_{x^2-y^2}$-wave gap has been observed 
in the $C(\phi)$ of CeCoIn$_5$~\cite{An2010PRL} and CeIrIn$_5$~\cite{Kittaka2012PRB} in the wide $T$ and $H$ regions.

Contrary to the expectation, however, 
no angular variation was detected within an experimental error in the $C(\phi)$ of UPt$_3$ under an in-plane rotation of $H$ ($\theta=90^\circ$) at temperatures $50 \le T \le 200$~mK. 
The $T$-$H$ ranges in which we searched for the $C(\phi)$ oscillation at $\theta=90^\circ$ are summarized in Fig.~\ref{HT} with squares.
For instance, the $C(\phi)$ data measured in the normal state at 88~mK and 2.8~T, and that measured in the SC state at 50~mK and 1.5~T
are shown in Figs.~\ref{phidep}(a) and \ref{phidep}(b), respectively. 
The $C(\phi)$ data measured under other $T$-$H$ conditions are also presented elsewhere.~\cite{Kittaka2012JPCS} 

On the basis of our fits to the data using the function $C_0(T)+C_H(T,H)(1-A_n\cos n\phi)$ ($n=2$ or 4), 
where $C_0$ and $C_H$ are the zero-field and field-dependent components of the electronic specific heat, respectively, and
$A_4$ ($A_2$) is the amplitude of the fourfold (twofold) oscillation normalized by $C_H$, 
$A_n$ never becomes larger than 0.3\% in the present $T$ and $H$ regimes.
This is in sharp contrast to the observation of a twofold oscillation in the $\kappa(\phi)$ of UPt$_3$ at 50~mK, 
whose amplitude is about 3\% with respect to the normal-state value $\kappa_{\rm n}$~\cite{Machida2012PRL}
as well as a fourfold oscillation in $C(\phi)$ with $A_4$ larger than 3\% at $0.4 \Tc$
in the $d_{x^2-y^2}$-wave superconductor CeIrIn$_5$ ($\Tc=0.4$~K).~\cite{Kittaka2012PRB}

If the SC gap has a horizontal line node at $k_z=0$ (e.g., the $E_{2u}$ model),
the anisotropy of $C(\phi)$ could be suppressed significantly at $\theta=90^\circ$ 
because $H$ is always in the nodal direction for $\theta=90^\circ$ [see Fig.~4(e) of ref.~\ref{Kittaka2012PRB}]. 
To examine this possibility, we measured $C(\phi)$ under a conically rotating $H$ titled by $60^\circ$ from the $c$ axis ($\theta=60^\circ$).
The $T$ and $H$ conditions under which we investigated $C(\phi)$ at $\theta=60^\circ$ are denoted in Fig.~\ref{HT} by circles.
Nevertheless, as presented in Fig.~\ref{phidep}(c), no angular variation was detected in $C(\phi, \theta=60^\circ)$ at 88~mK in $0.5 \le \mu_0H \le 2.2$~T.
These findings lead to a conclusion that 
QP excitations detected from the $C(\phi)$ measurements hardly depend on the field orientation around the $c$ axis.
We also observed that $C(\phi)$ measured after zero-field cooling and that after field cooling in 4~T applied above 10~K along one of the three equivalent $a$ axes 
provided the same results.

\begin{figure}
\begin{center}
\includegraphics[width=3.2in]{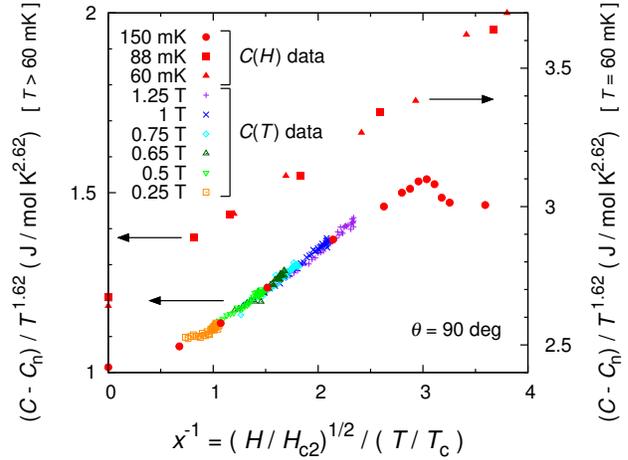}
\end{center}
\caption{
(Color Online) 
$(C-C_{\rm n})$ divided by $T^{1.62}$ at $\theta=90^\circ$ as a function of $x^{-1}=\sqrt{H/\Hc2}/(T/\Tc)$ in the intermediate-$T$ regime ($150 \le T \le 220$~mK),
where $\Tc=0.52$~K and $\mu_0\Hc2=2.6$~T.
The $C(H)$ data in the low-$H$ regime obtained at 60 and 88~mK are also plotted for comparison. 
}
\label{scaling}
\end{figure}

For nodal superconductors, QP DOS is expected to be scaled using a function of a single parameter $x=(T/\Tc)/\sqrt{H/\Hc2}$~\cite{Kopnin1996JETPL,Simon1997PRL,Volovik1997PRL}. 
Theoretical calculations predicted the scaling relations of various physical quantities, 
such as $C(H,T)/T^2 \propto f(x)$ and $\kappa(H,T)/T^3 \propto g(x)$ at low $T$,
where $f(x)$ and $g(x)$ are constant for a large $x$ and proportional to $1/x$ for a small $x$.
Indeed, $C(H,0) \propto \sqrt{H/\Hc2}$ and $C(0,T) \propto T^2$, which are expected for nodal superconductors, can be given by this scaling relation.
For UPt$_3$, it was reported that $\kappa/T^{2.7}$ in $H \parallel ab$ can be scaled by $x$ at low $H$ and low $T$.~\cite{Suderow1998PRL}
We found that $C(T,H)$ divided by $T^{1.62}$ can also be scaled by $x$ in the intermediate-$T$ regime.
Figure~\ref{scaling} shows $C/T^{1.62}$ as a function of $1/x$ plotted using the $C(T)$ and $C(H)$ data at $\theta=90^\circ$.
In $150 \le T \le 220$~mK, $C/T^{1.62}$ is proportional to $1/x$ when $x$ is small and tends to saturate when $x$ is large.
These features support that the specific heat of UPt$_3$ reflects QP DOS excited around the nodes.

However, at lower $T$ below 88~mK, $C/T^{1.62}$ does not follow the scaling relation owing to the unusual increase in $C(T)$ at low $T$. 
Although one might consider that the variation in $C$ by $H$, $\phi$, and $\theta$ is dominated by this unusual phenomenon,
it should be noted that $C(H)$ at 60 and 88~mK still shows the $\sqrt{H}$-linear dependence at low~$H$ (see Fig.~\ref{scaling}), and 
$C$ in the normal state is known to be robust to the intensity of $H$ up to 7~T down to 50~mK.~\cite{Schuberth1990EPL,Schuberth1995ZPB}
These findings imply that QP excitations still play an important role in determining the change in $C(H,\phi,\theta)$ at low $T$.

In order to settle the issue of the SC gap structure of UPt$_3$, 
the apparent discrepancy between the absence of any $C(\phi)$ oscillation and the presence of a twofold oscillation in $\kappa(\phi)$ should be resolved. 
A similar discrepancy has been reported on a filled skutterudite superconductor PrOs$_4$Sb$_{12}$ as well.\cite{Sakakibara2007JPSJ,Izawa2003PRL}
The discrepancies might be explained by considering the multi-band effect because 
$C(H,\phi,\theta)$ mainly detects the contribution of a heavy-mass band whereas $\kappa(H,\phi,\theta)$ may not. 
Indeed, UPt$_3$ has four heavy bands and one relatively light band.\cite{McMullan2008NJP}
In this context, the results of both $C(\phi)$ and $\kappa(\phi)$ experiments 
become consistent with each other if vertical line nodes exist only in the light band.
Such assumptions, however, seem to be very speculative at present because of the lack of experimental and theoretical justifications. 
Further efforts are needed to solve this issue.

\section{Summary}
We have measured the specific heat of UPt$_3$ at low temperatures down to 50~mK in various field orientations.
Under a magnetic field rotated in the $ac$ plane, 
the specific heat exhibits a large twofold oscillation at low temperatures and low fields with a deep minimum at $\theta=0^\circ$ ($H \parallel c$),
possibly reflecting the anisotropy of the intrinsic orbital limit of $\Hc2$. 
With increasing field or temperature, a striking peak develops at $\theta=0^\circ$ with a width of $|\theta| \lesssim 30^\circ$.
Correspondingly, at low $T$, an abnormal increase was observed in $C(H)$ for $|\theta| \le 30^\circ$.
Although its origin remains unclear, it might give the conditions in which the unusual $\Hc2$ limit becomes prominent in UPt$_3$.
Under a magnetic field rotated around the $c$ axis in $50 \le T \le 200$~mK, the specific heat hardly depends on the field orientation. 
Further experimental and theoretical investigations are essential to settle the SC gap structure of UPt$_3$
as well as the origin of the anomalous phenomena found for $|\theta| \lesssim 30^\circ$.

\acknowledgments
We thank K. Izawa, Y. Machida, and I. Vekhter for valuable discussions.
This work has partly been  supported by Grants-in-Aid 
for Scientific Research on Innovative Areas ``Heavy Electrons'' (20102002, 20102007, and 23102705)
from the Ministry of Education, Culture, Sports, Science and Technology of Japan and 
by Grants-in-Aid for Scientific Research (21340103) and Research Activity Start-up (22840013) 
from the Japan Society for the Promotion of Science (JSPS).

\end{document}